\documentstyle[epsf,aps,prl,multicol]{revtex}
\begin{document}

\begin{multicols}{2}

\noindent {\large\bf Comment on ``New Class of Resonances at the Edge of
the Two-Dimensional Electron Gas.''} \vspace{0.3cm}

Recently Zhitenev {\it et al.} (ZBAM) \cite{Zhitnev} reported 
capacitance, $C_g$, measurements between a surface metal gate and a
2DEG in the quantum Hall (QH) regime, 
revealing, as a function of gate voltage $U_g$, 
resonant transport between electrons propagating along
the sample edge and a ``puddle'' of electrons at higher filling factor isolated
under the gate. The frequency $f$ and magnetic field $B$ dependence 
of $C_g$ were similar for gates of widely varying size,
suggesting a common origin.

We recently employed 3D self-consistent electronic structure calculations
to elucidate the results of a similar experiment where, in a quantum
dot, backscattering of an asymptotically free edge state
through edge states trapped in the dot exhibited resonances as 
a function of $B$ and gate voltage $V_g$ \cite{us}. Both
experiments involve resonant scattering across an insulating, incompressible
strip (IS) between propagating and trapped edge states. Experience
from our experiment as well as a similar calculation for the
device in \cite{Zhitnev} (see figure) show unambiguously the 
following \cite{model}.

First, the resonant structure arises from an interference effect at
the bend in the free edge state
as it passes under (and out of) the gate. The area 
associated with the ``kink'' evolves with $U_g$ giving resonant
transmission when an integer number of flux quanta are enclosed.
Similar phenomena in point contacts
are well known.
For example, for $B=2.3 \; T$, we find
$dw/dU_g \sim 3 \; a_B^*/V$ where $w$ is the IS width and $a_B^*$ is
the effective Bohr radius. This gives a gate voltage spacing 
$\Delta U_g \sim 40 \; mV$, consistent with figure 2, in \cite{Zhitnev}.
This resonant resistance at the gate edges is trivially 
independent of gate length.
Also, consistent with the data, there can be
no resonance at $U_g=0$. Chklovskii \cite{Chklov} has recently
postulated that the resonance results from Coulomb regulated
tunneling through a ``dot'' embedded in the IS. Aside from the
necessarily random character of this mechanism, we find this
suggestion untenable because (a) the necessary dot size is
too large to fit in the IS and (b) there is no explanation for the
absence of resonances at $U_g=0$ (even for $B$ such that an IS exists). 

Second, as $U_g$ increases the edge states move to within
one or two magnetic lengths and, we postulate,
the resonant structure gives way to 
a total breakdown of the inter-edge state resistance and the
higher $C_g$ value is regained. 

Third, additional, frequency
dependent resistance, occurs {\it within the puddle}. Contrary
to the assumption of ZBAM, the most highly conductive region of 
the puddle is the perimeter, where the potential is still curving
(see in particular the $300 \; mV$ panel in figure).
The region (``about $10$ times larger'') which charges more
slowly is the compressible center, where the potential is very
flat, states are slightly below the Fermi energy $E_F$ and conductivity is low. 
Potential curvature increases with
$T$ explaining the {\it rise} in conductivity with $T$. Scaling
of the peaks with gate length also results from this 
resistance within the puddle. On resonance only this internal resistance
produces frequency dependence; off resonance the weak tunneling 
between edges contributes a second frequency dependent part.

\begin{figure}[hbt]
\setlength{\parindent}{0.0in}
 \begin{minipage}{\linewidth}
\epsfxsize=8cm
\epsfbox{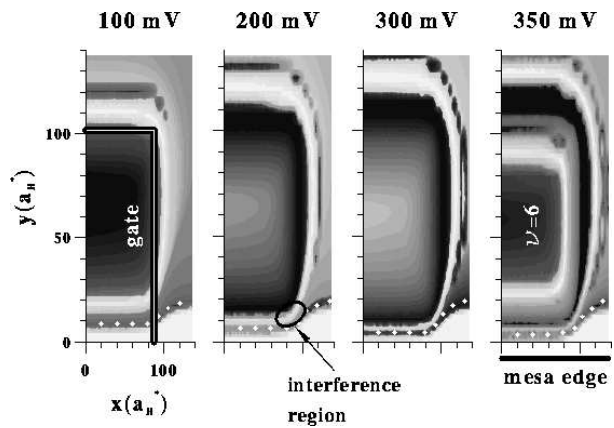}
\vspace*{3mm}
\caption{Density of states within $\pm 0.25 \; K$ of $E_F$ for device 
of [1] at $B=2.3 \; T$. Mesa edge (shown in right panel)
along $x$-axis ($y \approx -10 \; a_B^*$),
gate {\it width} is $1.1 \; \mu m \approx 110 \; a_B^*$, gate edge shown
in left panel. Light bands are IS, propagating $\nu=1,2$ edge state 
(highlighted with dots) bends as $U_g$ increases; oval in second
panel denotes interference region. $\nu=6$ terrace
forms suddenly and $C_g$ drops at $U_g \sim 350 \; mV$, but $\nu=4$
terrace remains connected so drop is not so great.}
 \end{minipage}
\end{figure}

Finally, calculations including disorder are possible, but for
a high mobility device we expect the conclusions to be 
qualitatively unaffected. Small random changes in the kink dimensions can
be expected accounting for shifts in the exact peak structure with 
thermal cycling. \vspace{0.3cm}

\noindent M. Stopa and J. P. Bird \\
Riken \\
Saitama, Japan \vspace{0.3cm}


\end{multicols}


\begin{references}
\bibitem{model} The condition 
$dU_g/dB =  \nu e/C_g h c$, is simply the condition for constant 
filling factor $\nu$. We use this and the slope of the leading
edge of the capacitance drop in 
Fig. 2 of \cite{Zhitnev} to estimate the effective
gate to 2DEG distance as $\sim 160 \; nm$.
\end{references}
\end{document}